\documentclass{emulateapj} 

\usepackage{times}

\newcommand{\etal}{{et al.}}
\newcommand{\eg}{{\it e.g.,}}
\newcommand{\ie}{{\it i.e.,}}

\shorttitle{Extended X-ray Emission from QSOs} 
\shortauthors{Stockton et al.} 
\journalinfo{accepted for publication in ApJ}
\submitted{accepted for publication in ApJ}

\begin{document}

\title{Extended X-Ray Emission from QSOs\footnotemark[1]}

\footnotetext[1]{Based in part on data  obtained with the Chandra X-Ray Observatory, which is 
operated for the National Aeronautics and Space Administration by the Smithsonian
Astrophysical Observatory. Also based in part on observations made with the NASA/ESA 
Hubble Space Telescope, obtained from the Data Archive at the Space Telescope 
Science Institute, which is operated by the Association of Universities for Research 
in Astronomy, Inc., under NASA contract NAS 5-26555.}

\author{Alan Stockton, Hai Fu, and J. Patrick Henry}
\affil{Institute for Astronomy, University of Hawaii, 2680 Woodlawn
 Drive, Honolulu, HI 96822}
 
 \and

\author{Gabriela Canalizo}
\affil{Department of Physics and Institute of Geophysics and Planetary Physics, 
University of California, Riverside, CA 95521}

\begin{abstract}
We report Chandra ACIS observations of the fields of 4 QSOs showing
strong extended optical emission-line regions.  Two of these show no
evidence for significant extended X-ray emission. The remaining two fields,
those of 3C\,249.1 and 4C\,37.43, show discrete (but resolved) X-ray
sources at distances ranging from $\sim10$ to $\sim40$ kpc from the
nucleus.  In addition,
4C\,37.43 also may show a region of diffuse X-ray emission extending
out to $\sim65$ kpc and centered on the QSO.  

It has been suggested that extended emission-line regions such as
these may originate in the cooling of a hot intragroup medium.  We do
not detect a general extended medium in any of our fields,
and the upper limits we can place on its presence indicate cooling times
of at least a few $10^9$ years.

The discrete X-ray emission sources we detect cannot be explained as
the X-ray jets frequently seen associated with radio-loud quasars, nor
can they be due to electron scattering of nuclear emission.  The most
plausible explanation is that they result from high-speed shocks from
galactic superwinds resulting either from a starburst in the QSO host
galaxy or from the activation of the QSO itself. Evidence from densities
and velocities found from studies of the extended optical emission 
around QSOs also supports this interpretation.

\end{abstract}


\section{Introduction}

\begin{figure*}[!t]
\epsscale{0.945}
\plotone{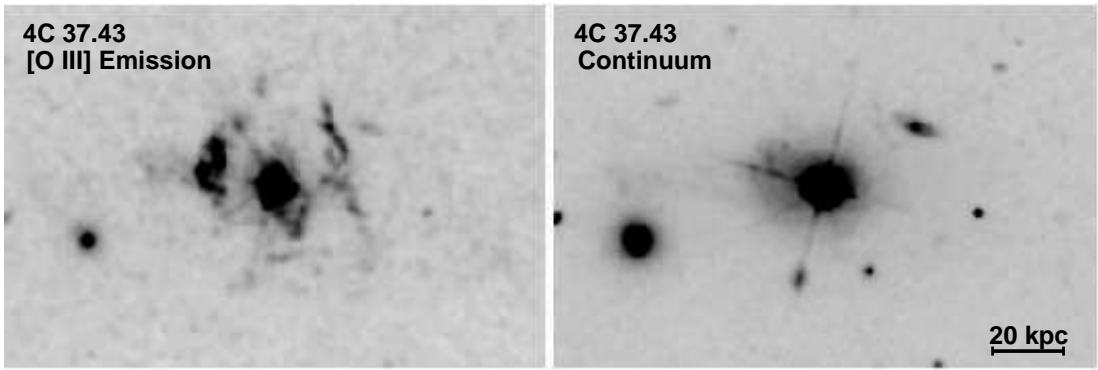}
\caption{\footnotesize HST WFPC2 images of the $z=0.37$ quasar 4C\,37.43.  The image on the
left is a narrow-band linear-ramp-filter image centered on the [O\,{\sc iii}]
$\lambda5007$ line.  The image on the right is a broad-band F814W image,
dominated by continuum light. (H$\alpha$ falling within the edge of this
filter profile is responsible for the very weak presence of the strongest
emission features). See Stockton \etal\ 2002 for more details.  In this and all
following images, north is up and east is to the left.}\label{4c37img}
\end{figure*}

A significant fraction of low-redshift QSOs show strong extended emission
over radii from the nucleus of a few tens of kiloparsecs or more \citep{wam75,sto76,
ric77,bor85,sto87}.  Briefly, the main features of these extended emission-line regions
(EELRs) are that (1) they are often highly structured, with several discrete
knots and/or filaments, (2) their morphologies generally do not follow that
of either the QSO host galaxy or the radio emission, (3) the inferred ionization
parameter indicates that at least some of the emission comes from relatively
dense regions ($\sim$ a few hundred cm$^{-3}$), and (4) the most luminous EELRs
are associated with steep spectrum radio-loud QSOs.  Currently, the best studied of 
these EELRs is that of the $z=0.37$ quasar 4C\,37.43 \citep{sto02}, shown in 
Fig.~\ref{4c37img}. The origin of the gas comprising these EELRs remains uncertain. 
Currently, there are three main scenarios:
\begin{enumerate}
\item The gas is tidal debris from an interaction or merger, photoionized by
UV radiation from the nucleus \citep{sto87}.
\item The gas comes from a cooling flow. The warm gas condenses from hot ($10^8$ K) 
gas in a halo surrounding the quasar and is, again, photoionized by the nucleus \citep{fab87}.
\item The gas has been blown out and shocked, either by a superwind from a vigorous 
starburst or by a wind generated by the turning on of the QSO (\eg\ \citealt{diM05}). 
The gas itself may originally have been either tidal debris (as in 1) or simply ambient gas 
associated with the host galaxy. The gas is photoionized by the UV photons from the 
nucleus and/or ionized locally by thermal emission from the high-speed shocks propagating 
through the gas \citep{sto02}.
\end{enumerate}
Crawford, Fabian, and their colleagues have insisted quite strongly that
EELRs around QSOs and radio galaxies must have their origins in cooling flows
\citep{fab87,cra88,cra89,for90,cra00}.
They have pointed out that emission regions with the inferred densities will 
dissipate quite rapidly unless confined by some mechanism, and they suggest 
that the most reasonable confinement mechanism is a hot external medium.
To support this interpretation, they have measured the [\ion{O}{2}]/[\ion{O}{3}]
line ratio and used photoionization models to estimate the ionization parameter
and thence the pressure in the EELR. Assuming pressure equilibrium with a hot
external medium allows an estimate of the cooling time for the latter. In most cases,
this time is found to be much shorter than the Hubble time.

However, there has been no {\it direct} evidence that cooling flows are
important in these cases, and, in any case, the standard cooling-flow paradigm
seems to have been ruled out by recent observations \citep[\eg][]{pet03}.  
Furthermore, \citet{sto02} have shown from detailed
photoionization modeling of the EELR around the quasar 4C\,37.43 that at 
least two density regimes are required, and that these cannot be in pressure 
equilibrium with each other.  
This model indicates that a problem with using the [{O}\,{\sc ii}]/[{O}\,{\sc iii}] 
line ratio
to infer the ionization parameter is that most of the [{O}\,{\sc ii}] emission
comes from regions with a very small ($\sim10^{-5}$) filling factor and 
densities of several hundred cm$^{-3}$, while almost all of the [{O}\,{\sc iii}] 
emission comes from regions with essentially unity filling factor and 
densities $\sim200$ times lower. The temperatures of these two regions
are found to be comparable. The lower-density 
component could plausibly be in pressure equilibrium with a hot external 
medium; the denser component cannot be.  The confinement mechanism 
for the dense regions has to be something 
other than hydrostatic pressure (most likely
transient shocks), and the pressure of any hot gas component has to be at least
two orders of magnitude lower than that found by \citet{cra00}. The cooling
time for such a gas is longer than a Hubble time. This conclusion, of course,
depends on whether the best-fitting photoionization model is roughly close
to reality; while it was selected from an extensive grid of over 200 model runs, 
there remain the usual worries about the uniqueness of the solution. 

In order to further explore the nature of these EELRs, and, in particular, to test
whether a hot gaseous halo of sufficient pressure to have a short cooling time
is present, we have obtained deep imaging of three fields with the ACIS on board 
the Chandra X-ray Observatory. In this paper, we discuss the results of this imaging
program, including as well ACIS imaging of one additional QSO EELR field from the
Chandra archive.  In \S\ref{obs}, we describe the observations and reduction
procedures.  In \S\ref{analysis}, we discuss the morphologies of the extended
X-ray components, comparing them with the optical extended emission.  Finally, in
\S\ref{discussion}, we attempt to place these X-ray observations in the overall 
context of possible models for EELRs. We assume a flat cosmological
model with $H_0=70$ km s$^{-1}$, $\Omega_m=0.3$, and $\Omega_{\Lambda}
= 0.7$.

\section{Observations, Reduction Procedures, and Basic Results}\label{obs}

Recently, four QSOs having luminous EELRs have been observed with the ACIS onboard 
the {\it Chandra} X-ray Observatory (3 from our program no.\ 04700368 and 1 from
program no.\ 04700768, retrieved from the {\it Chandra} archive).  We give the relevant parameters
for these observations in Table \ref{obsparam}. 
\begin{deluxetable}{l c c c c c}
\tablewidth{0pt}
\tablecaption{Chandra ACIS Observations of QSOs with EELRs}
\tablehead{
\colhead{QSO} & \colhead{Redshift} & \colhead{ACIS CCD} & \colhead{Exposure}
& \colhead{Seq No.} & \colhead{Obs ID} 
}

\startdata
MRK\,1014 & 0.159 & S3 & 12 ks & 700783 & 4104 \\
3C\,249.1 & 0.312 & I2 & 24 ks & 700665 & 3986 \\
4C\,25.40 & 0.250 & I2 & 16 ks & 700666 & 3987 \\
4C\,37.43 & 0.371 & I2 & 42 ks & 700667 & 3988 \\
\enddata
\label{obsparam}
\end{deluxetable}

Instead of using the level 2 event files produced by the standard pipeline
calibrations, we have gone back to the level 1 event files. The main purpose
in doing this was to
recover valid source events that were incorrectly rejected as afterglow
events by the standard data
processing. New level 2 event files were generated after applying the Good Time
Intervals (GTIs) to clean the data of time intervals with high
background levels. The data reduction was done with CIAO 3.1 (Chandra Interactive
Analysis of Observations)\footnote{http://cxc.harvard.edu/ciao/}.

We computed model PSFs for each QSO using the Chanda Ray Tracer tool (ChaRT;
\citealt{car03}),
which carries out a detailed, frequency-dependent, ray trace of
the Chandra optics.  We extracted the spectrum of each QSO, using only the
grade 0 events (which should be essentially free of pile-up concerns for
the flux levels of our sources), and used this model spectrum as input
to ChaRT, but with the exposure times increased so that we get roughly
the same number of counts as in the ``all good grade'' images.
The simulated distributions of rays were then projected onto the ACIS-I
or ACIS-S detector plane, as appropriate, and generated as event files using
MARX\footnote{http://space.mit.edu/CXC/MARX/}.

As explained in \S~\ref{analysis}, we cannot achieve reliable PSF subtraction
within $\sim2\arcsec$ radius of the QSOs.  Beyond this radius, for two of these fields, 
those of Mrk\,1014 and 4C\,25.40, we detect no significant extended component at all. 
For the remaining two, 4C\,37.43 and 3C\,249.1, extended
emission is observed, but it does not have the character of a hot gaseous halo
that would give evidence for a cooling flow. Instead, the X-ray emission appears
to be highly structured, and it may to some extent
follow the distribution of some of the $\sim10^4$ K gas producing the 
extended optical emission. The X-ray emission does not
closely follow the radio structure, so this extended component is quite
different from the X-ray jets often found around quasars. The remainder of
this paper focuses on this extended X-ray emission.

\section{The Nature of the Extended X-Ray Emission}\label{analysis}
\subsection{The Morphologies of the Extended X-Ray Components}
In order to detect and analyze any extended X-ray components, we need to achieve a
fairly accurate subtraction of the PSF. For Chandra images, this PSF removal is 
complicated by two effects: (1) The Chandra PSF is dependent on the X-ray energy, so it is
necessary to model the PSF as a weighted sum of mono-energetic PSFs, and (2) regions
of high X-ray flux suffer from {\it pile-up}, such that, if two X-rays fall within the sample time
(the default 3.2 s, in our case) and within a $3\times3$-pixel detection box, the event may
either be rejected or interpreted as a single detection of an X-ray at a higher energy. Pile-up
thus results both in loss of counts and in the distortion of the spectral energy distribution.
While pile-up should be relatively moderate for our sources, and completely negligible
for regions beyond about a 2\arcsec\ radius, it could cause problems for our PSF subtraction.
We have therefore adopted the following procedure.  We extract an image from the level 2
event file over the energy band from 0.5 to 7 keV, using only grade 0 (single pixel) events.
These are much less subject to pile-up than are the events of other grades that are 
normally folded in to make the standard pipeline level-2 image.
We model the spectrum of this grade-0 image as a power law combined with photoelectric 
absorption, and use this model spectrum as the input to ChaRT for producing PSF ray traces.
Using MARX, we generate 100 quasi-independent realizations of the PSF and average
these to obtain a low-noise PSF. We scale this PSF to the observed level-2 image at
a radius of 2 arcsec, where pileup should be negligible, yet the flux is still likely to be dominated
by the QSO point source.  We then subtract this scaled PSF from the level 2 image, 
ignoring the region interior to a 2\arcsec\ radius, as well as any residuals that clearly are due
to slight mismatches between the shapes of the QSO PSF and the synthetic PSF at radii 
where the profile has a steep gradient.

\begin{figure*}
\epsscale{1.0}
\plotone{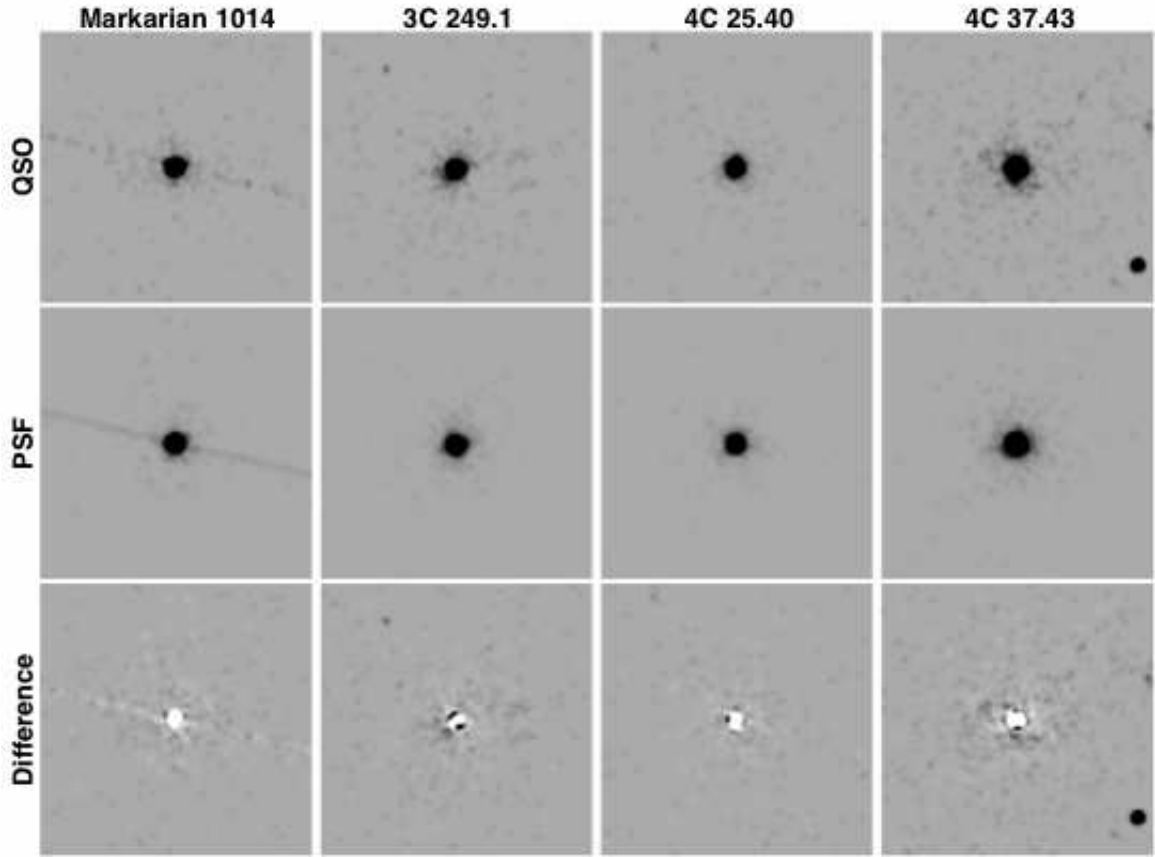}
\caption{\footnotesize Background-subtracted Chandra ACIS images, scaled synthetic PSFs, 
and differences for
the four fields discussed in this paper. All images have been smoothed with Gaussians
with $\sigma=1$ pixel, giving a point-source FWHM of $\sim1\farcs8$. The fields 
shown are 1\arcmin\ square. North is up and East to the left.}\label{psfsub}
\end{figure*}

\begin{figure*}
\epsscale{0.6}
\plotone{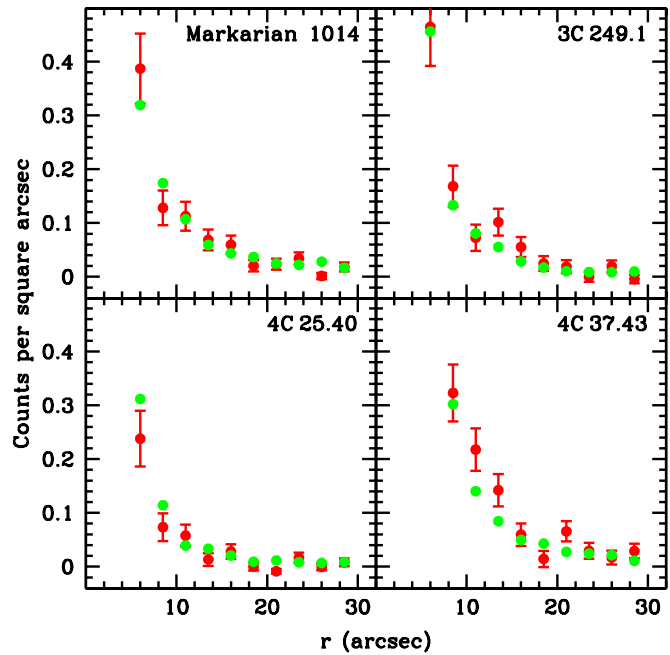}
\caption{\footnotesize Outer X-ray radial-surface-brightness profiles for the targets in our
sample ({\it red}) and our scaled synthetic PSFs ({\it green}). The random errors for the latter
are approximately the size of the points or smaller.}\label{radpsf}
\end{figure*}

Figure \ref{psfsub} shows the results of this subtraction for each of the four fields, and 
Fig.~\ref{radpsf} shows the outer parts of the radial-surface-brightness profiles, compared
with those of the corresponding scaled synthetic PSFs. We evaluate the significance of
apparent features in the difference images from photon statistics based on the counts
in the observed images and the scaled PSF models.
We discuss the fields individually in the following sections.
\subsubsection{Markarian 1014}
Markarian 1014 is both the lowest redshift and the only radio-quiet QSO in the sample. It shows
the strongest optical extended emission of any of the radio-quiet QSOs in the sample of
\citet{sto87}, but this emission is still more than a factor of 10 less luminous than the
strongest extended emission found in steep-spectrum radio-loud QSOs. The host galaxy
is clearly the result of a recent merger: it is an ultraluminous IR source (\eg\ \citealt{san88}), 
and the optical
image shows an enormous, broad, tidal ``arm'', with an even larger, much lower surface
brightness, counter-arm (\eg\ \citealt{mac84,can00}).

The ACIS image shows no significant evidence for extended X-ray emission, either as discrete 
sources (Fig.~\ref{psfsub}) or as a diffuse extended component (Fig.~\ref{radpsf}).

\subsubsection{3C\,249.1}
The extended optical emission around 3C\,249.1 was first noticed spectroscopically by
\citet{ric77} and imaged by \citet{sto83,sto87}. There appears to be no deep imaging of
the host galaxy itself in bands that are not contaminated by the extended emission.

The synthetic PSF is not a perfect match to the actual PSF, as indicated by the large residuals
to the NE and SW in the subtracted image (Fig.~\ref{psfsub}). Nevertheless, there is clear
evidence for discrete clumps of X-ray emission to the east and southeast at a radius of about 3\arcsec.
There may possibly be some additional clumps to the west at $\sim15\arcsec$ radius, but 
Fig.~\ref{radpsf} shows no evidence for a general diffuse component.
\begin{figure*}
\epsscale{0.5}
\plotone{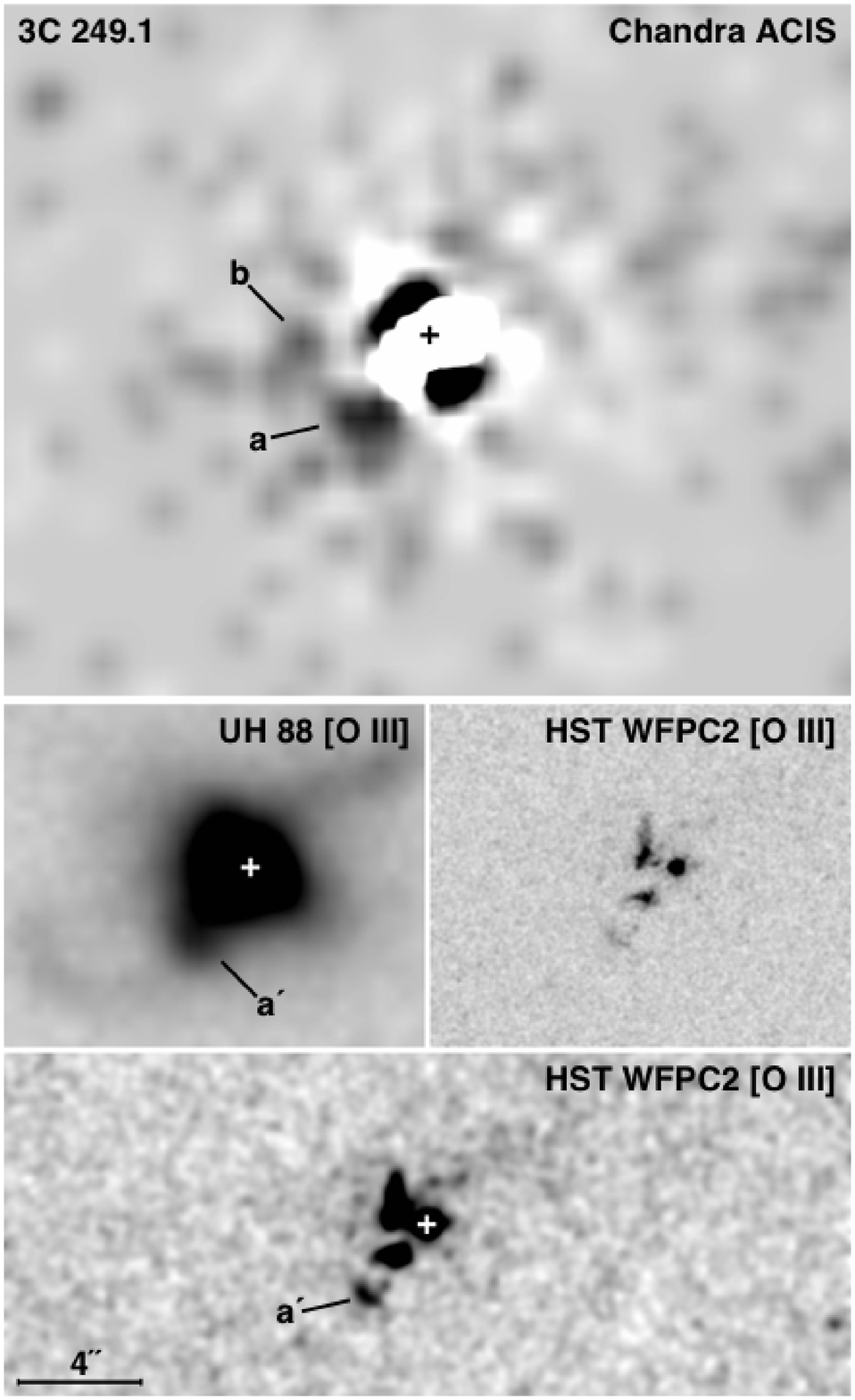}
\caption{\footnotesize\ Chandra ACIS-I and [\ion{O}{3}] images of 3C\,249.1. The
upper panel shows a higher-contrast version of the difference image from
Fig.~\ref{psfsub}. Two significant regions of off-nuclear X-ray emission are labeled.
(Note that, both here and in Fig.~\ref{4c37.43mos}, the hard stretch and a zero level
set just below the sky level tends to overemphasize the negative values just beyond
a 2\arcsec\ radius, due to the PSF subtraction in regions where there are no X-ray
events in the image.)
The lower panel shows narrow-band images centered on the redshifted 
[\ion{O}{3}] $\lambda5007$ emission line. In the upper left of the lower panel is shown 
a ground-based image of 3C\,249.1,
taken through a 30 \AA\ filter (from \citealt{sto83}; see also \citealt{sto87}).  
The remaining two subpanels show, at different contrasts, a [\ion{O}{3}] image obtained 
through the F656N filter with PC1 channel of WFPC2 on {\it HST}. All images are at 
the same scale, given in the lower-left corner of the lower panel. Crosses indicate
the position of the quasar in cases where this location may be uncertain.} \label{3c249.1mos}
\end{figure*}
A comparison of the ACIS image for 3C\,249.1 with optical emission-line
images is shown in Fig.~\ref{3c249.1mos}. The ground-based [\ion{O}{3}] image is
reproduced from \citet{sto83}, and the short-exposure {\it HST} WFPC2 image was 
retrieved from the {\it HST} archive.  The brightest detected region of X-ray emission, 
3\arcsec\ SE of the nucleus (labeled $a$), is nearly coincident with a region of optical 
emission, seen as 
a small, arc-like structure in the {\it HST} image (labeled $a'$). Any X-ray emission coincident 
with the 
higher-surface-brightness regions of optical emission closer in would have been swamped 
by the nuclear X-ray source. There is no detected [\ion{O}{3}] emission at the position of 
the weaker X-ray source $b$. With a proper intensity stretch, both $a$ and $b$ can be seen 
clearly on the ACIS images prior to PSF subtraction.  Their significance is dominated by
photon statistics rather than by any uncertainty in the PSF subtraction. Object 
$a$ is detected at $3 \sigma$, while object $b$ is detected at just over $2 \sigma$.

\subsubsection{4C\,25.40}
This quasar has a highly structured, filamentary extended emission region in the optical lines
(see Fig.~2$j$ in \citealt{sto87}).  In our Chandra ACIS image, however, there is no evidence
for any significant extended emission either as discrete clumps or as a general surrounding
medium.

\subsubsection{4C\,37.43}
4C\,37.43 has the most luminous extended optical emission-line region of any known
for low-redshift ($z<0.5$) QSOs. The host galaxy is clearly detected in both ground-based
and {\it HST} line-free continuum images (\citealt{sto02} and references therein; see also
Fig.~\ref{4c37img}); the surface-brightness distribution is asymmetrical, being enhanced
towards the east, and there is faint, thin filament extending in the same direction at least 60 kpc 
from the southern side of the host.
\begin{figure*}
\epsscale{0.5}
\plotone{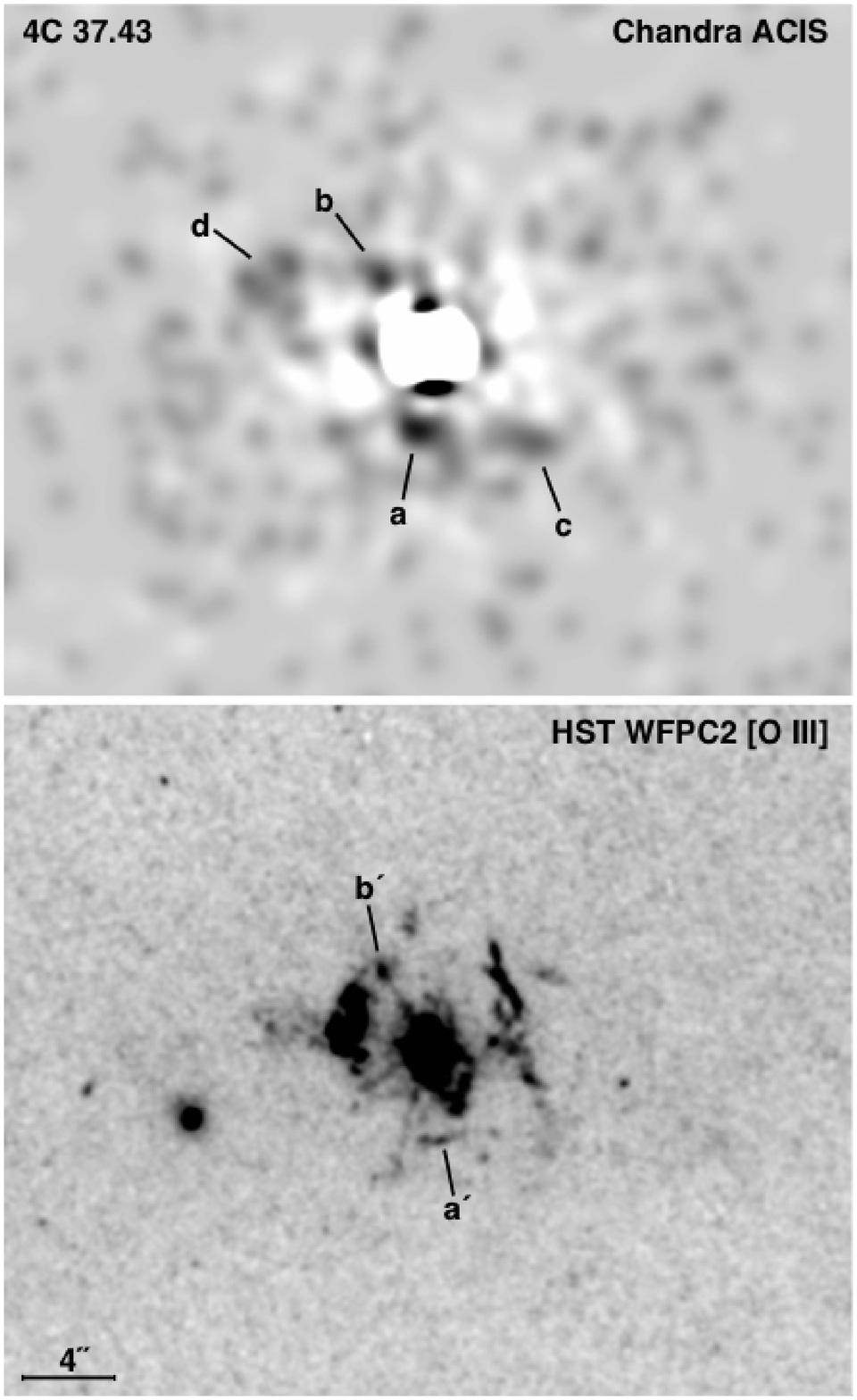}
\caption{\footnotesize\ Chandra ACIS-I and [\ion{O}{3}] images of 4C\,37.43. The
upper panel shows a higher-contrast version of the difference image from
Fig.~\ref{psfsub}. Several discrete off-nuclear X-ray emission sources are labeled. 
The lower panel shows a deep {\it HST} WFPC2 image obtained through
a linear ramp filter centered on the redshifted 
[\ion{O}{3}] $\lambda5007$ emission line \citep{sto02}. The sources labeled
$a'$ and $b'$ may be associated with the corresponding X-ray sources.
Both images are at 
the same scale, given in the lower-left corner of the lower panel.} \label{4c37.43mos}
\end{figure*}
The X-ray emission shows several discrete sources at radii ranging from 4\arcsec\
to 13\arcsec\ (Fig.~\ref{4c37.43mos}), one of the most luminous being a clump (labeled $d$
in Fig.~\ref{4c37.43mos}, detected at $\sim4 \sigma$) at a radius of about 8\arcsec\ 
(40 kpc, projected) 
to the east-northeast.  This is in the same direction as the strongest optical emission, but at about
twice the radius.  There is weak optical emission extending out to just south of this position.
Other strong clumps of X-ray emission are found to the south, northeast and 
southwest (labeled $a$, $b$,
and $c$, respectively); $a$ and $b$ have discrete [\ion{O}{3}] sources just beyond the
X-ray emission.  There is no discrete  [\ion{O}{3}] source associated with $c$,
although it falls in a region of low-surface-brightness diffuse emission. Object $a$ is
detected at just over $3 \sigma$; objects $b$ and $c$ are each over $2.5 \sigma$.
Weaker clumps of X-ray emission are found in an slightly elliptical halo, out to about 13\arcsec\ 
radius.  These are apparently responsible for the $\sim2 \sigma$ enhancement  
of the X-ray radial-surface-brightness profile over
the PSF model seen between 10\arcsec\ and 14\arcsec\ in Fig.~\ref{radpsf}. Beyond 14\arcsec,
however, the observed and model PSF profiles are essentially identical, once again indicating
the lack of a significant extended X-ray halo component.

\subsection{Upper Limits on Extended Diffuse X-Ray Emission}

Galaxy groups often show diffuse extended X-ray emission from a hot intragroup
medium, indicating the existence of a common potential well sufficiently deep to
retain such gas.  We have pointed out in our discussions of each of our four fields
that we do not see evidence for a diffuse, very extended component.  We now need to
quantify these statements in terms of upper limits to the presence of such gas.
Because of the small numbers of detected events we have to work with, we cannot
use constraints from the expected radial surface-brightness profile of the extended
component.  Instead, we simply deal with counts in a defined annulus, assuming
a reasonable model for the distribution of the emission.

We use the annulus from 10\arcsec\ to 30\arcsec. A 10\arcsec\ inner radius is sufficient
to eliminate most of the wings from the QSO PSF, and, at these redshifts, probably most
of any diffuse X-ray emission specifically associated with the host galaxy (although,
as described below, 4C\,37.43 is an exception). The 30\arcsec\
outer radius keeps us within a typical group core radius, even for 4C\,37.43, the highest
redshift object in our sample.

For definiteness, we assume a so-called $\beta$ (modified King) model 
\citep{cav78,bir93} for the surface brightness distribution:

\begin{math}
\sigma(r) = \sigma_0[1 + (r/r_c)^2]^{-3\beta\ +\ 0.5},
\end{math}

\noindent
where $\sigma_0$ is the central surface brightness and $r_c$ is the core radius. The 
corresponding density distribution is given by

\begin{math}
n = n_0[1 + (r/r_c)^2]^{-3\beta/2}.
\end{math}

\noindent
Basing our choice on the two-component fits to group X-ray emission by \citet{mul98},
we assume typical parameters for the extended intragroup medium of $r_c = 175$ kpc,
$\beta=1$, and $kT=1$ keV.  This value for $\beta$ for the extended component
is steeper than that found from single-component fits, which \citet{mul98} found
to give unacceptable residuals in most cases.

We determine the counts in the annulus both for the background-subtracted ACIS
observations of the four fields and for the corresponding scaled synthetic PSF
images.  From the photon statistics, we determine the standard deviation of
the difference in the annulus, and we take 3 times this value as a conservative
upper limit to any contribution from an extended X-ray component. In all cases
except the 4C\,37.43 field, the difference between the counts in the annulus
for the observation and for the synthetic image were $<1\ \sigma$; for 4C\,37.43,
the counts for the observation are 2 $\sigma$ above those for the PSF model.
These extra counts are, however, entirely due to the previously mentioned excess
associated with the host.  If the inner radius of the annulus is increased from 10\arcsec\
to 14\arcsec, the excess in the annulus completely disappears.

From the counts corresponding to the $3\ \sigma$ upper limits in the annulus in
each of the fields, we normalize $\beta$ models with the above parameters at
the appropriate redshift.  From these models, we calculate the counts we would
have expected to receive over the whole circular area out to $10r_c$.  After converting
to count rates, we then use 
WebPIMMS\footnote{http://heasarc.gsfc.nasa.gov/Tools/w3pimms.html} 
to calculate the corresponding luminosity, assuming a Raymond-Smith 
model with 0.4-solar metallicity.

In order to obtain a rough upper limit to the central
density or pressure, we need the cooling coefficient $L(T)$, where
the volume emissivity $\Lambda(T)=L(T)n_{e}n_{p}$.  At $kT=1$ keV ($T\approx1.2\times10^7$
K), line emission is an important contributor to the cooling. We use the calculation of
the cooling coefficient given by \citet{sut93}, which gives $L=1.48\times10^{-23}$
erg cm$^3$ s$-1$. We then integrate the emissivity over the
volume out to $r_c$ to find the value of $n_{p0}$ that gives the luminosity found with
PIMMS.  For these upper limits to the central densities, the lower limits to the cooling times
for the gas at the center ranges from $t_c > 7.4\times10^9$ years for the 4C\,37.43 
field to $t_c>14.2\times10^{9}$ years for the MRK\,1014 field. All of the observed and 
calculated parameters are given in Table \ref{upperlimit}. While some of the lower limits
to the cooling time are formally slightly less than a Hubble time, they do not offer much 
encouragement to the
view that strong cooling is taking place in these groups.  Furthermore, as \citet{sut93} stress,
these estimates, which assume equilibrium conditions and no effect from the radiation
field, will give cooling times considerably lower than those for (more plausible)
non-equilibrium models. 
\begin{deluxetable}{l c c c c c}
\tablewidth{0pt}
\tablecaption{Upper Limits to Diffuse Extended Emission}
\tablehead{
\colhead{QSO} & \colhead{Annular} & \colhead{Model} & \colhead{Luminosity}
& \colhead{$n_{p0}$} & $t_c$\\
& Counts\tablenotemark{a} & Counts\tablenotemark{b} & (erg s$^{-1})$ & (cm$^{-3}$)
& (Gyr)
}
\startdata
MRK\,1014 & 30 & 92 & $<1.4\times10^{43}$ & $<0.0014$ & $>14.2$ \\
3C\,249.1 & 25 & 58 & $<4.2\times10^{43}$ & $<0.0024$ & \phn$> 8.3$\\
4C\,25.40 & 24 & 53 & $<3.3\times10^{43}$ & $<0.0022$ & \phn$> 9.0$\\
4C\,37.43 & 45 & 76 & $<5.2\times10^{43}$ & $<0.0027$ & \phn$>7.4$ \\
\enddata
\tablenotetext{a}{Counts corresponding to 3 times the $\sigma$ of the difference between
the signal in the annulus from 10\arcsec\ to 30\arcsec\ radius for the background-subtracted
Chandra ACIS image and that for the scaled model PSF image.}
\tablenotetext{b}{Total counts within a radius corresponding to $r_c=175$ kpc for a $\beta$ model
(see text for details) that has been scaled to the counts in column 2 within an annulus from
10\arcsec\ to 30\arcsec\ radius.}
\label{upperlimit}
\end{deluxetable}

\section{The Origin of the X-ray Emission}\label{discussion}

In the two fields for which we detect extended sources of X-ray emission, these sources
have distributions similar in extent to those of the extended optical emission.  In some
cases there may even be correlations between the optical and X-ray
brightness peaks; in other cases there clearly is no such correlation.  We consider here
various possible emission mechanisms for the X-rays.

\begin{enumerate}

\item The most common extended X-ray features seen around quasars are X-ray jets
\citep[see, \eg][and references therein]{sam04,mar05}.
These correspond closely in position and morphology to radio jets and are evidently
due either to Compton upconversion of microwave background photons by relativistic electrons
in the jets or to X-ray synchrotron emission. 3C\,249.1 shows a radio jet on the same side 
as the extended X-ray emission
\citep{mil93,kel94}, but it is at position angle $\sim99\arcdeg$, placing it {\it between}
our objects $a$ and $b$. While no evidence for a radio jet in 4C\,37.43 has yet been detected
\citep{mil93}, it is a classic FR2 double source, and none of the X-ray emission 
components lie along the radio axis.  In neither case do X-ray jets provide a satisfactory
explanation for the observed extended X-ray emission.
\item Some extended X-ray emission in principle might be due to electron
scattering of the nuclear X-rays, as has been suggested for some of the near-nuclear soft
X-ray emission from the Cygnus A radio galaxy \citep{you02}. 
However, the low Thompson cross-section makes electron scattering 
an extremely inefficient process, so it is useful to estimate the actual scattered flux in a
typical example. While we do not have direct information on the electron densities in the
X-ray emitting regions, they are unlikely to be greater than those in the most luminous
optical EELRs.  We have the required information for the
extended emission region about 4\arcsec\ east of 4C\,37.43. The total mass of the gas in this EELR is 
$\sim$4 M$_{\odot}$ (Stockton et al. 2002), which gives a total number of electrons of 
$\sim 10^{57}$, assuming fully ionized pure hydrogen. A typical distance for the X-ray
sources is $\sim$20 kpc 
from the nucleus, and we assume that the nuclear X-ray flux in their direction is the
same as towards our line of sight. The calculation gives a ratio of scattered to
direct nuclear flux of $\sim10^{-14}$ for a region of typical size. Since, for 4C\,37.43, we 
detect $\sim6700$ counts from the nucleus in our 
42 ks of integration, we would expect $<10^{-10}$ counts due to scattering in regions like
those we observe. Electron
scattering thus fails rather dramatically in this context, even if some of our assumptions
were to be incorrect by a few orders of magnitude. 
\item In some cases \citep{sak00,kin02}, the X-ray emission can be dominated by X-ray
recombination lines from the $\sim10^4$ K photoionized gas. Our calculations based on
photoionization models for the strongest optical emission region associated with 4C\,37.43 
\citep{sto02} indicate that this mechanism could potentially make a significant contribution,
but also that there are many uncertainties.  The fact that we do not actually detect X-ray
emission from any of the most luminous optical extended emission regions in 4C\,37.43,
nor from any of the extended emission regions of MRK\,1014 or 4C\,25.40 argues
against its importance generally in the cases we are considering.
\item X-ray emission can be thermal emission and line emission from high-speed 
shocks. The cases where there {\it may} be a spatial correlation between the optical
emission-line gas and the X-ray gas (i.e., 3C\,249.1$a$ and 4C\,37.43$a$ and $b$) 
are reminiscent of the classical picture 
of galactic superwinds. It has been shown in many edge-on starburst galaxies (\eg NGC\,253, 
NGC\,4945) that the X-ray gas shows a limb-brightened filamentary structure, and lies either 
coincident with, or just to the inside of, H$\alpha$ emission-line filaments \citep{cecil02,str04}. 
The underlying mechanism is most likely either that (1) as the high-speed hot wind and the cold ISM 
intermingle, the dense ambient gas becomes entrained and heated in the wind fluid, resulting in
both optical emission and local increases in the X-ray emissivity; or (2) as the superwind encounters 
a large cloud, it produces a slow ($10^2$ km s$^{-1}$) radiative shock into the cloud and a fast 
($10^3$ km s$^{-1}$) stand-off bow shock; the former produces the optical emission, the latter 
the X-rays \citep{leh99}.
Similarly, in a case where the extended X-ray emission does tend to follow the extended 
optical emission, this correlation
may be due to thermal emission from shocks within the extended emission region,
produced by superwinds associated with a recent interaction- or merger-driven starburst. 
These shocks would also provide a natural explanation for the higher-density
regions found from modeling the optical emission-line spectrum.
\end{enumerate}

If a galactic superwind due to a starburst is indeed the cause of the extended optical and X-ray 
emission around 
quasars, then there are at least two predictions that can be used to test the model:
(1) There must be high-velocity shocks in the emission-line regions, and the thermal emission 
from these shocks may, in some regions, dominate the photoionization of the surrounding gas
by UV radiation from the quasar nucleus; 
(2) The post-starburst ages of the host galaxies must be young, \ie\ on the order of tens of Myr or 
less. The time it takes for the high-speed wind gas to travel from the nucleus to the EELRs is less 
than a few tens of Myr, given a typical wind velocity of $10^3$ km s$^{-1}$ and a typical EELR 
distance of a few tens of kpc from the center of the host galaxy. 

In order to carry out the first test, two conditions must hold:  (1) there must be regions where
shocks are present but the gas is shielded from a direct line-of-sight to the nucleus, and
(2) we must be able to distinguish the signature of ionization from
shocks from that of photoionization by the quasar continuum.  This latter requirement is complicated by the fact
that much of the emission-line radiation from a high-velocity shock can be due to photoionization of the
pre-shock gas by the thermal continuum from the shock itself. Such
``autoionizing'' shock models are distinguished from nuclear photoionization models by the shape 
of the ionizing-photon spectrum in the vicinity of the emission-line regions. There are intrinsic
differences, with the nuclear continuum following an approximate power law over many decades in
frequency, in contrast to the thermal spectrum of the shock. 
But the nuclear emission is also likely to be subject to higher photoelectric absorption of
soft X-rays due to intervening material, further increasing the relative hardness ratio of the ionizing 
radiation.  These differences in the ionizing continuum are reflected in the ratios of certain 
diagnostic emission lines, and both of these forms of photoionization can be distinguished 
from each other and from pure shock ionization by their emission spectra \citep{vil97,all98}.  Unfortunately, 
the most useful strong lines tend to be in the ultraviolet region inaccessible from the ground for 
objects of modest redshift. However, with efficient spectrographs on large ground-based telescopes,
it is now feasible to use weaker lines, such as [\ion{O}{3}] $\lambda4363$ and \ion{He}{2} $\lambda4686$. 
\citet{eva99} have shown that the 
[O\,{\sc iii}] $\lambda4363$/[O\,{\sc iii}] $\lambda5007$ vs. He\,{\sc ii} $\lambda4686$/H$\beta$ 
diagram could be a very powerful diagnostic for discriminating nuclear photoionization models 
and high-speed shock models (see their Fig.~10). Realistic nuclear photoionization models, 
shock photoionization models, and pure shock-ionized models are all reasonably well separated 
in this diagram.

We have so far been assuming that a global superwind can only be due to a starburst.
However, another possibility is that the superwind could be a direct result of the turning on
of the quasar itself \citep{diM05}. In this case, although the test for whether shock models
can explain the line emission is still appropriate, we wouldn't necessarily expect to see
an extremely recent starburst.

Whatever the origin of the high-velocity gas, the detection of discrete X-ray features
associated with two of our quasar fields seems clearly to point to the presence of
high-speed shocks. This picture is consistent with evidence from extended optical
emission, in particular, (1) the existence of small regions of dense gas that cannot
be in hydrostatic pressure equilibrium with their surroundings, and (2) gas velocities
that exceed those that can reasonably be attributed to gravitation. Given the lack of
correlation of any of these features with radio jets, some sort of galactic wind seems
the most reasonable possibility.

\acknowledgments
Support for this work was provided by the National Aeronautics and Space
Administration (NASA) through Chandra Award Number GO3-4126 issued by the
Chandra X-ray Observatory Center, which is operated by the Smithsonian Astrophysical
Observatory for and on behalf of NASA under contract NAS8-03060.
This research has also been partially supported by NSF grant AST03-07335. 
It made use of the NASA/IPAC Extragalactic Database (NED) 
which is operated by the Jet Propulsion Laboratory, California Institute of 
Technology, under contract with the National Aeronautics and Space 
Administration. The authors recognize the very significant
cultural role that the summit of Mauna Kea has within the indigenous
Hawaiian community and are grateful to have had the opportunity to
conduct observations from it.


\end{document}